Submitted to Nuclear Fusion.
IAEA FEC 2021 contribution CN 286

# Experimental confirmation of efficient island divertor operation and successful neoclassical transport optimization in Wendelstein 7-X

Thomas Sunn Pedersen[1,2], for the W7-X Team*
[1]Max Planck Institute for Plasma Physics, Garching and Greifswald, Germany
[2]University of Greifswald, Greifswald, Germany
*See team list in Appendix 1

## Abstract

*We present recent highlights from the most recent operation phases of Wendelstein 7-X, the most advanced stellarator in the world. Stable detachment with good particle exhaust, low impurity content, and energy confinement times exceeding 100 ms, have been maintained for tens of seconds. Pellet fueling allows for plasma phases with reduced ITG turbulence, and during such phases, the overall confinement is so good (energy confinement times often exceeding 200 ms) that the attained density and temperature profiles would not have been possible in less optimized devices, since they would have had neoclassical transport losses exceeding the heating applied in W7-X. This provides proof that the reduction of neoclassical transport through magnetic field optimization is successful. W7-X plasmas generally show good impurity screening and high plasma purity, but there is evidence of longer impurity confinement times during turbulence-suppressed phases.*

## 1. Introduction

Wendelstein 7-X (W7-X), which has a magnetic field strength of 2.5 T and a plasma volume of 30 $m^3$, started operation in 2015 [1-5]. Its mission is manifold, but can be broadly summarized as delivering experimental proof that the optimized stellarator concept is a viable and attractive concept for a fusion power plant, and that the computational optimization approach taken can be successful, despite the complex and rich physics phenomena observed in fusion plasmas. In contrast to most other magnetic confinement fusion concepts being pursued, the confinement and stability of a stellarator plasma is primarily dictated by the details of the vacuum magnetic field geometry and topology, and less by self-consistent plasma effects. In particular, the stellarator does not rely on driven currents parallel to the vacuum magnetic field for confinement. This makes a stellarator, once built, less flexible than a tokamak, but all the more computationally tractable. The computational optimization approach for the stellarator requires that the magnetic topology be created from the external coils with high accuracy to keep error fields small. In stellarators, error fields can be measured (through the use of flux surface mapping), and corrected (eg. using relatively simple trim coils) to a very high accuracy, as was done in W7-X before first plasma operation [6].

The lack of driven parallel currents gives the stellarator its well-known and well-established advantages of steady-state capability, a lack of runaway electron damage during disruptions, and, at the power-plant stage, significantly lower recirculating power. W7-X is a particularly clear example of the computational approach and the relatively small effects of parallel plasma currents, since it was optimized to minimize plasma-generated parallel equilibrium currents, ie. the bootstrap- and Pfirsch-Schlüter currents. The successful minimization of bootstrap current was experimentally demonstrated early in operation [5]. Two further – and particularly important -experimental demonstrations of the W7-X optimization will be summarized here: The successful reduction of neoclassical transport [7], and efficient operation of the island divertor [8, 9].

Important insights gained with respect to turbulent transport, and the role of impurities will also be presented. These represent examples of physics issues that are less well understood than eg. neoclassical transport, and are therefore more challenging to optimize for. These days, computational tools exist that can address issues such as turbulence suppression [10], but such tools were not available, or were only in their nascence, at the time that the W7-X design was computationally optimized two decades ago.

The results reported here were obtained in operation phases 1.2a in 2017, and 1.2b in 2018, which were performed after the installation of a full set of in-vessel components, in particular 10 passively cooled fine-grain graphite test divertor units, and upgrades to the heating and fueling systems, including higher ECRH power, and the addition of NBI heating, pellet fueling, and density control using several different gas fueling systems [eg. 11]. Despite the lack of active cooling, plasma pulses up to 100 s were successfully sustained [12] at 2 MW of heating, and also plasmas exceeding 30 second duration at 5 MW of heating, of which more than 26 seconds were with divertor detachment [8, 9]. The device is now being upgraded again, this time with a full set of water-cooled plasma-facing components, divertor cryopumps, and extensions of fueling and heating systems. Once completed, the device will be commissioned and restart plasma operation in 2022, having reached the hardware capabilities allowing for the exploration of higher-performance, near-steady-state plasmas.

## 2. Divertor operation

During attached operation, large wetted areas were seen, in particular in the standard configuration for W7-X. These routinely exceeded 1 $m^2$, and showed a tendency to increase with heating power, giving hope that high-power operation will be safe even at high heating power levels with the water-cooled divertor currently being installed, whose specified maximum heat flux is 10 $MW/m^2$ [13]. This appears to be possible also for discharges exceeding the 10 MW plasma heating originally specified as the power level for quasi-steady-state operation). As reported earlier [14], the heat load patterns generally agree very well with numerical predictions, but there were some surprises which have largely been understood at this point [15, 16, 17].

The wetted area in a divertor can be thought of as a result of the balance in the scrape-off layer (SOL) between parallel transport (free-streaming towards the targets) and perpendicular transport. The long connection lengths in the W7-X SOL [18] should allow for a wide SOL in W7-X [19], but this width depends also on the radial transport. Using a combination of stationary, divertor-target-integrated Langmuir probes, and a reciprocating Langmuir probe at the outer midplane, experiments have been performed that shed light on transport in the W7-X SOL [20, 21]. Localized density perturbations, usually referred to as blobs, in W7-X are, as expected, highly elongated along the magnetic field. Their perpendicular motion is primarily poloidal, and does not show the strong radial (perpendicular to the outer flux surfaces) “ballistic” nature seen in tokamaks [22]. Nonetheless, the large wetted areas observed indicate that radial transport in the W7-X SOL is still substantial.

In a fusion reactor, the longevity of a divertor depends not only on the heat flux being low enough that the plasma-facing material temperatures can be kept below their melting or ablation points, but also that the plasma temperatures are low enough that physical sputtering, primarily caused by ions accelerated through the sheath potential, is negligible. Heat fluxes as well as plasma temperatures at the divertor targets are benign if divertor detachment can be achieved. In W7-X, stable and complete detachment was achieved routinely. The pumping efficiency was initially relatively low [14, 23] but it significantly improved later, after boronization allowed a reduction of the radiating mantle of oxygen and carbon, for reasons at least partly related to better divertor plugging [14], but other mechanisms have also been identified [24]. Detachment with high pumping efficiency was achieved for up to 26 seconds at a heating power of 5 MW with a very low impurity content [10] (Figure 1), indicating control of divertor-heat-flux, plasma density, and impurity content, and giving confidence for reaching the foreseen high-performance, quasi-steady-state (30 minutes) discharges in the future [25].

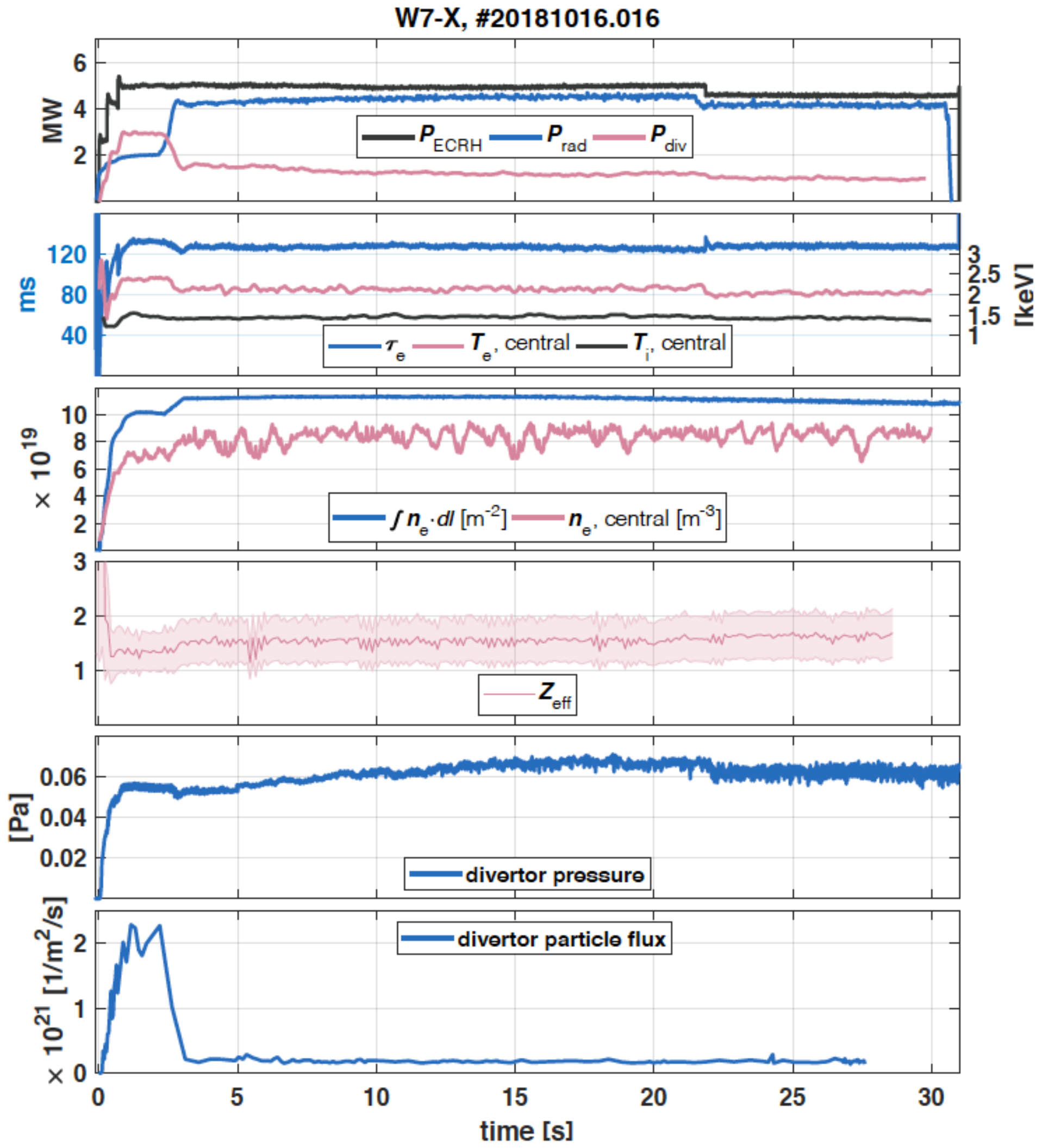

*Figure 1: Main parameters for the discharge that was stably detached with good particle exhaust, low impurity content, stable density and stored energy for more than 28 s, until the pre-programmed discharge end [10].*

The performance of the W7-X divertor, and the behavior and parameters of the edge- and scrape-off-layer plasma are now understood in quite some detail [eg. 26], thanks to measurements from a suite of diagnostics [eg. 27-30] as well as further modeling. The discharge of Fig.1 approached steady-state in terms of the overall particle inventory, including the substantial inventory of hydrogen in the mostly graphite-based plasma-facing components, and towards the end of the discharge, the measured rate of particle exhaust through the divertor turbopumps was almost equal to the actively injected particle rate, as seen in Fig. 2.

Detailed two-dimensional patterns of particle flow parallel to the magnetic field have now been measured in the SOL of W7-X using a Coherence Imaging Spectroscopy (CIS) diagnostic [27, 28, 31, 32]. Multiple counter-streaming $C^{2+}$ impurity flows are seen – and in most cases closely agree with numerical predictions. The experiments performed in the last operational campaign show that the $C^{2+}$ impurities are well-coupled to the flows of the main ions and generally stream toward the nearest divertor target, as measured along the magnetic field. CIS 2D measurements have also been performed in reversed-field experiments, showing measurable differences to forward-field flows. This indicates that perpendicular drifts – presumably ExB drift - should be included in future SOL modeling. Recent investigations [28] show that the particle parallel flow velocities are sensitive to electron density variations, and a significant drop in flow velocity is seen to be a clear indicator of the onset of detachment, and thus provides a second, independent signal, in addition to the infrared camera measurements of divertor heat flux, for detachment detection and control.

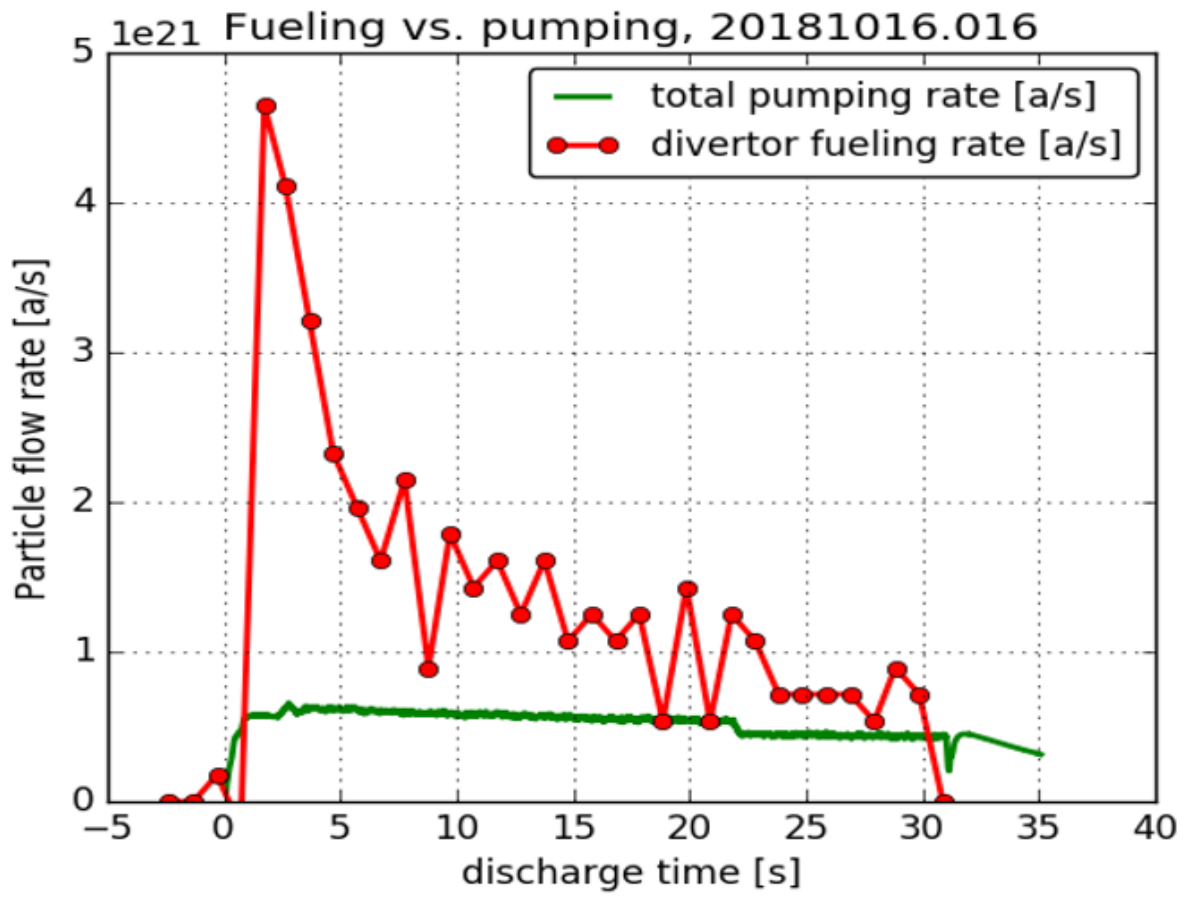

Figure 2. *For the detached discharge in Figure 1, the particle fueling rate (in red) was only slightly higher than the particle exhaust rate (green) towards the end of the discharge, indicating that a steady-state inventory was nearly reached, and that wall-absorption was playing a minor role.*

## 3. High-performance discharges and proof of neoclassical optimization

High-performance discharges, such as the earlier-reported stellarator triple-product record discharge which exceeded $6 \times 10^{19}$ keV $m^{-3}$ s [33], give us the opportunity to prove that the optimization for reduced neoclassical transport in W7-X was successful [7]. In Fig.3, we present one such high-performance discharge, which is comparable to the already-published triple-product-record discharge, with a very similar confinement time, but with a somewhat higher density, and somewhat lower ion temperature: For this discharge, the temperatures were appr. 2.5 keV for ions and 3 keV for electrons in the center, whereas the plasma density was $1 \times 10^{20}$ $m^{-3}$ in the core. The discharge was heated with 4.5 MW of heating, and had an energy confinement time of 0.23 s corresponding to about 1.2 times the energy confinement time expected from the ISS04 stellarator scaling [34]. Values at the ISS04 scaling or slightly above are broadly consistent with tokamak H-mode plasmas, as shown in Fig. 4 of ref. 5. For other, less optimized stellarators scaled to the W7-X size and magnetic field strength, similar plasma temperature and density profiles would have required significantly higher heating power to balance neoclassical transport, in particular in the mid-radius (strong gradient) region, as shown on the right-hand-side of the figure. High-performance discharges are generally characterized by core density peaking, and a reduction of turbulent density fluctuations. Without such turbulence reduction, the central ion temperature appears to be clamped to appr. 2 keV [35]. These findings are consistent with W7-X transport usually being dominated by ITG turbulence, but stabilized by strong density gradients in a so-called *stability valley* [36] as exemplified in Figure 4 for the W7-X standard configuration. The smaller linear growth rates in the *stability valley* have been shown in nonlinear gyrokinetic turbulence simulations to translate to lower saturated transport rates and are not due to radial electric field effects [37]**.**

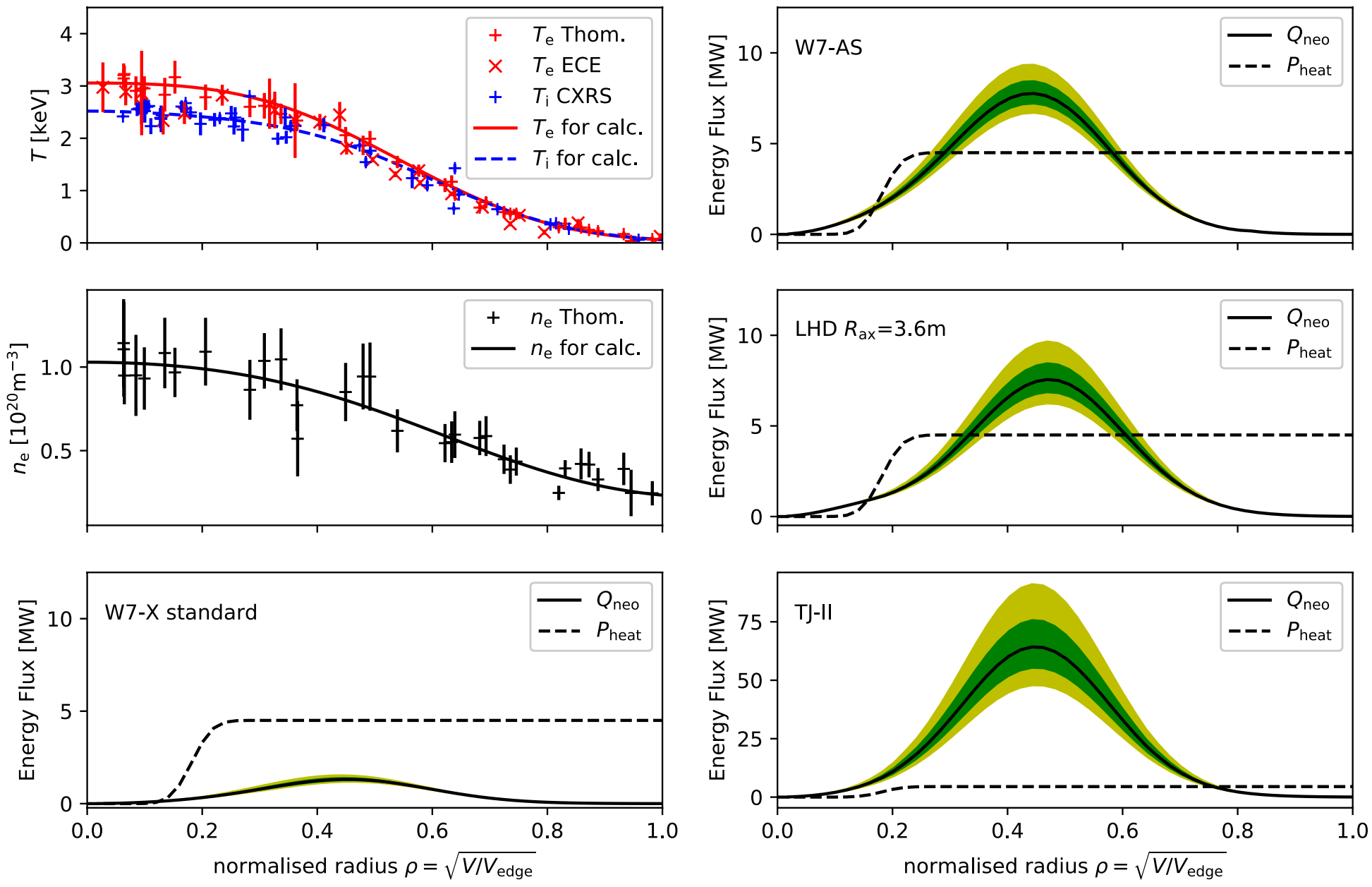

Figure 3 *Proof of successful reduction of neoclassical transport in W7-X: Left: The temperature (top) and density profiles (middle) in discharge 20180918.045 for the time window around t~3.35 s: For W7-X (bottom left), the neoclassical transport is shown with solid a solid black line, calculated from the measured density and temperature profiles, with error bars in green. It accounted for a relatively small fraction of the radial transport of the 4.5 MW of ECRH deposited in the central region. For less optimized devices (right panels), such plasma values would not have been achievable at 4.5 MW of heating, since the neoclassical transport loss rate would have exceeded the applied heating power significantly. Dotted lines: The heating power density integrated up to the given normalized radius.*

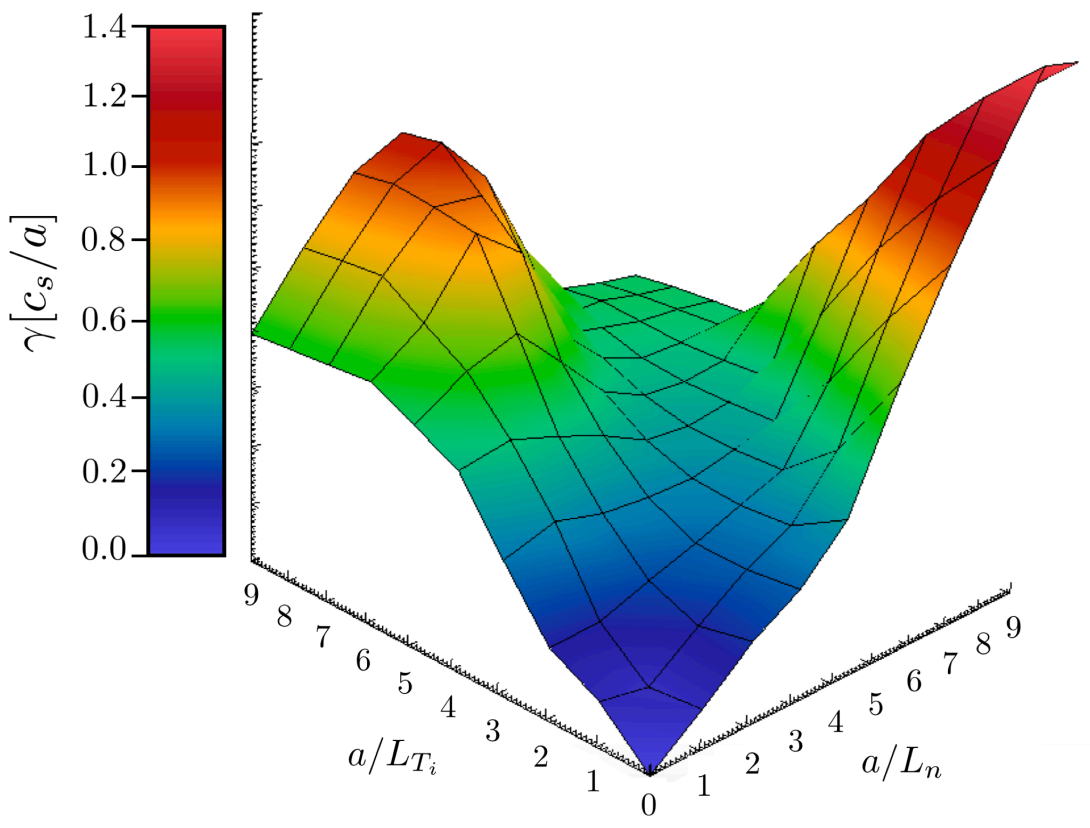

Figure 4: *The normalized linear growth rate of electrostatic modes, calculated by the GENE code for the W7-X standard configuration, shows a reduced growth rate in a "valley" where the temperature-*

*and density gradient scale lengths are comparable, suggesting that a significant temperature gradient can be sustained without causing strong turbulence, if the density gradient is also significant. This was indeed the case for the transient high-performance phase of the triple-product record shot. This calculation was done assuming $T_e=T_i$. See also [36].*

Although there are a number of examples of discharges, or discharge periods, with good performance (comparable to what is just described), and these show good reproducibility, the majority of discharges in W7-X were performed without pellet injection and show strong turbulent transport, and relatively low particle and energy confinement times, see Fig. 5. These are usually also characterized by electron temperatures being significantly larger than ion temperatures – related to the ion temperature clamping just mentioned. The difference between these two classes of discharges (well-confined with $T_e$ close to $T_i$, or poorly confined with $T_e$ greater than $T_i$), and the often rather fast transitions between these two states, lead one to think of them as bistable states and speculate to what degree the underlying dynamics may display hysteresis, as is seen in other situations, such as L-mode to H-mode dynamics. In the following we will present a heuristic explanation of why the discharges tend to end up in one or the other situation in W7-X. Many of the pieces of this puzzles are themselves solidly established and have been published in the literature.

We start with the sources for particles and energy in W7-X plasmas. Most discharges are heated primarily or exclusively by ECRH. This heating source heats exclusively the electrons, and the absorption is usually close to the plasma center. Most discharges are fueled with gas puffing from either the main gas inlet systems (with gas nozzles residing more than a meter away from the last closed flux surface), or the divertor gas inlet system (with gas nozzles residing just a few centimeters away from the last closed flux surface). With a central heat source, exclusively onto the electrons, and a peripheral particle source, it is to be expected that the density profiles are relatively flat, and that the electron temperature profile is centrally peaked. These profiles therefore bring the plasma close to the left "red shoulder" in Figure 3, where the ITG growth rate is substantial. One therefore expects high ITG turbulence, ie. large ion thermal losses and low particle confinement time. The latter perpetuates (or even strengthens) the flat density profiles, whereas the former leads to the ion temperatures to go down, or stay low. Although this reduces the ion temperature gradient (stabilizing for ITG) it also increases the Te/Ti ratio, which is destabilizing for ITG [35]. Therefore, the flattening and decrease of the ion temperature profiles is stronger than one would expect from Fig 4 alone.
The interconnectedness of various effects leading to relatively low confinement and depressed ion temperatures is illustrated in Figure 5 below.

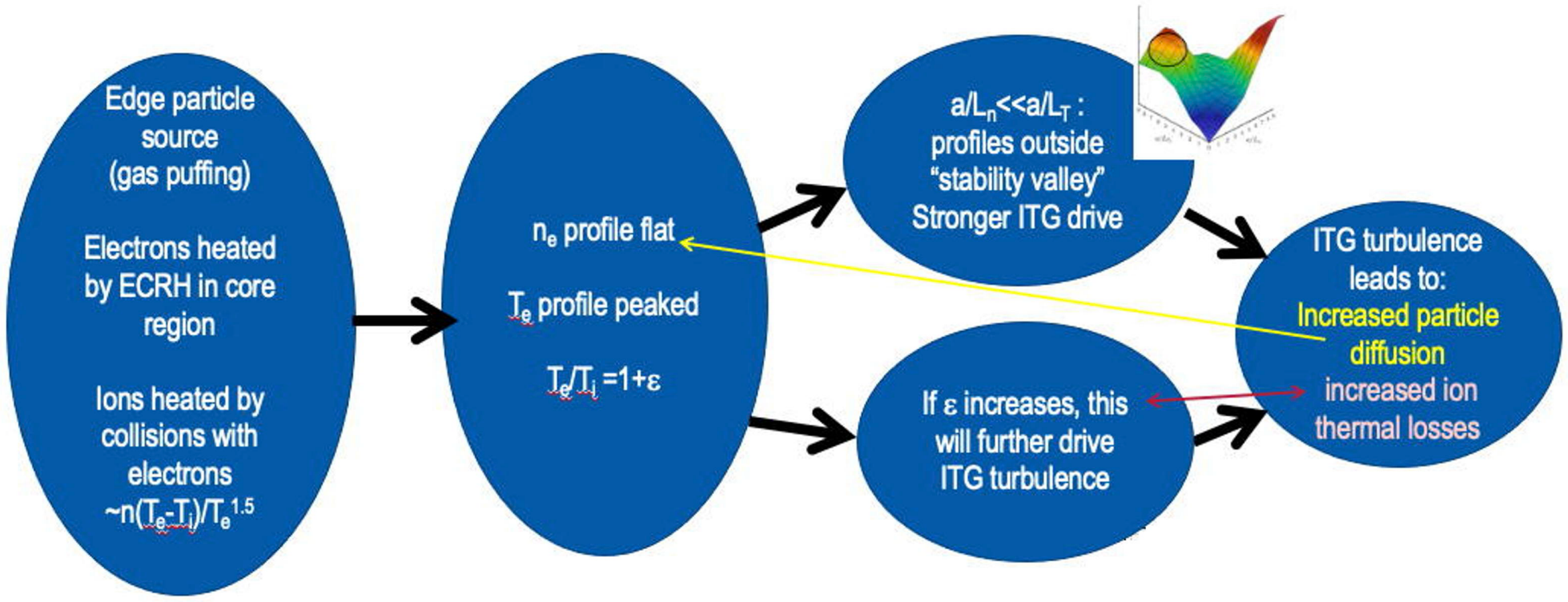

Figure 5. *A low-confinement "vicious cycle" is started by having central electron heating and peripheral particle fueling and is perpetuated because of details of ITG stability.*

A scatter plot of triple products for discharges without pellet injection is shown in Fig. 6 below. A general tendency for improved triple products at higher densities is seen, but performance is well below expectations based on neoclassical transport estimates. These discharges in general have relatively flat density profiles. One exception is NBI heated discharges which often display peaked density profiles. In Fig. 6, they do not stand out as having improved confinement, but it should be cautioned that their triple products are likely underestimated, since port losses, shine-through, and fast ion losses were not taken into account when calculating their confinement times. Neutral beam power deposition simulations are still undergoing validation [38, 39]. NBI results are discussed also in Section 5.
Recent progress in our understanding of the scaling of confinement with density can be found in Ref. [40] and will also be briefly discussed in Section 7.

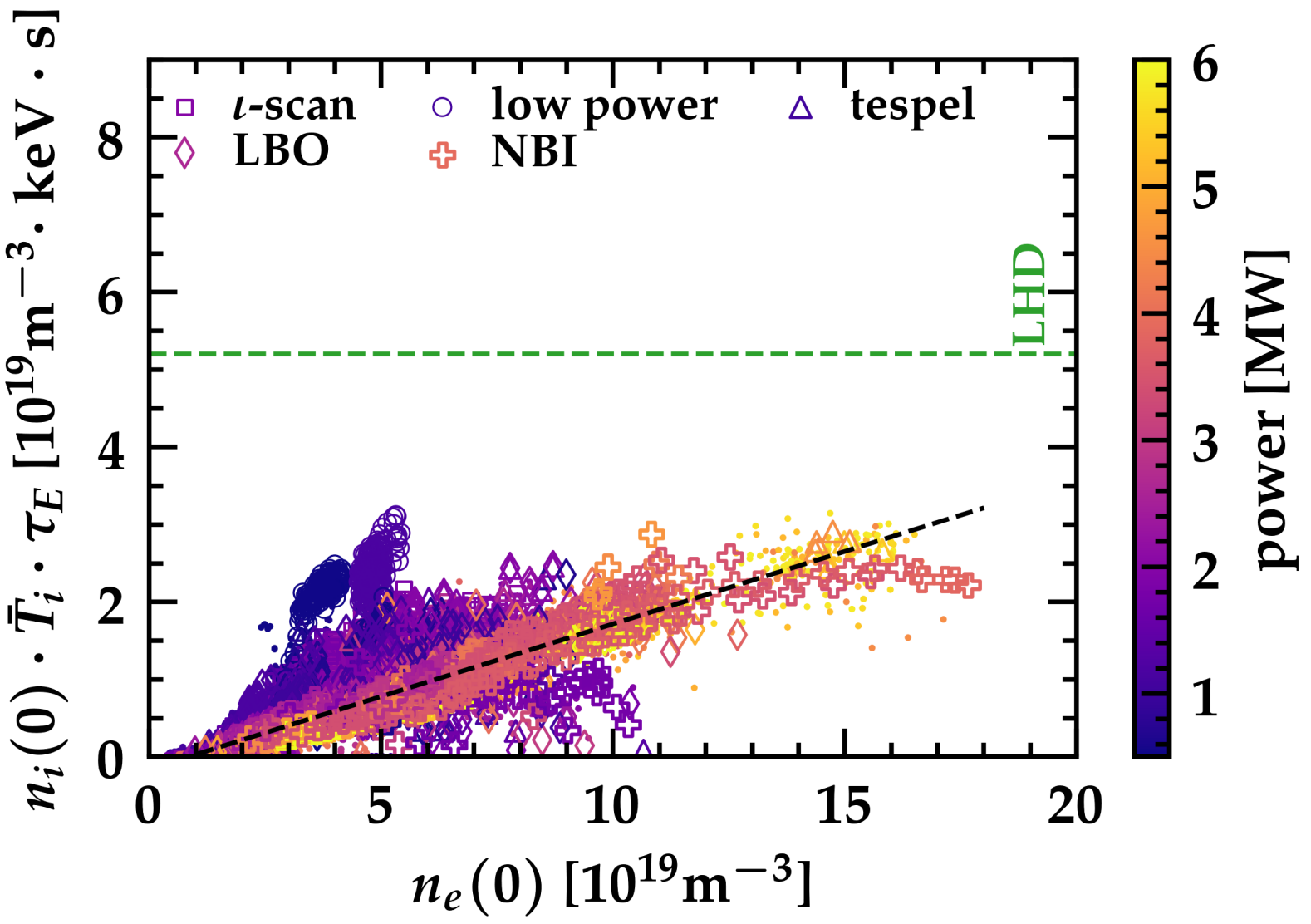

Figure 6. *Discharges without pellet injection have relatively modest triple products presumably due to the lack of density peaking.*

This "vicious cycle" can be broken if a powerful central particle source is added.

As the density begins to peak, the profiles move in to the "stability valley" on Fig. 4, the ITG turbulence reduces, the ion temperature can increase and approach the electron temperature and this further stabilizes the ITG. Moreover, with the significantly improved particle confinement (combined with the central particle source), the density peaking is more easily maintained. This allows for further ion temperature peaking without exiting the "stability valley". In OP1.2 the density peaking was achieved using the blower-gun pellet injection system [41], which, despite its shortcomings (relatively low pellet injection speed of 300 m/s, and only of order 40 pellets available per discharge), allowed for high-performance phases to be achieved on numerous occasions, albeit only with short duration (less than 1 sec). In Fig. 7 discharges with pellet injection have been added to the scatter plot from Figure 6, clearly showing their higher performance.

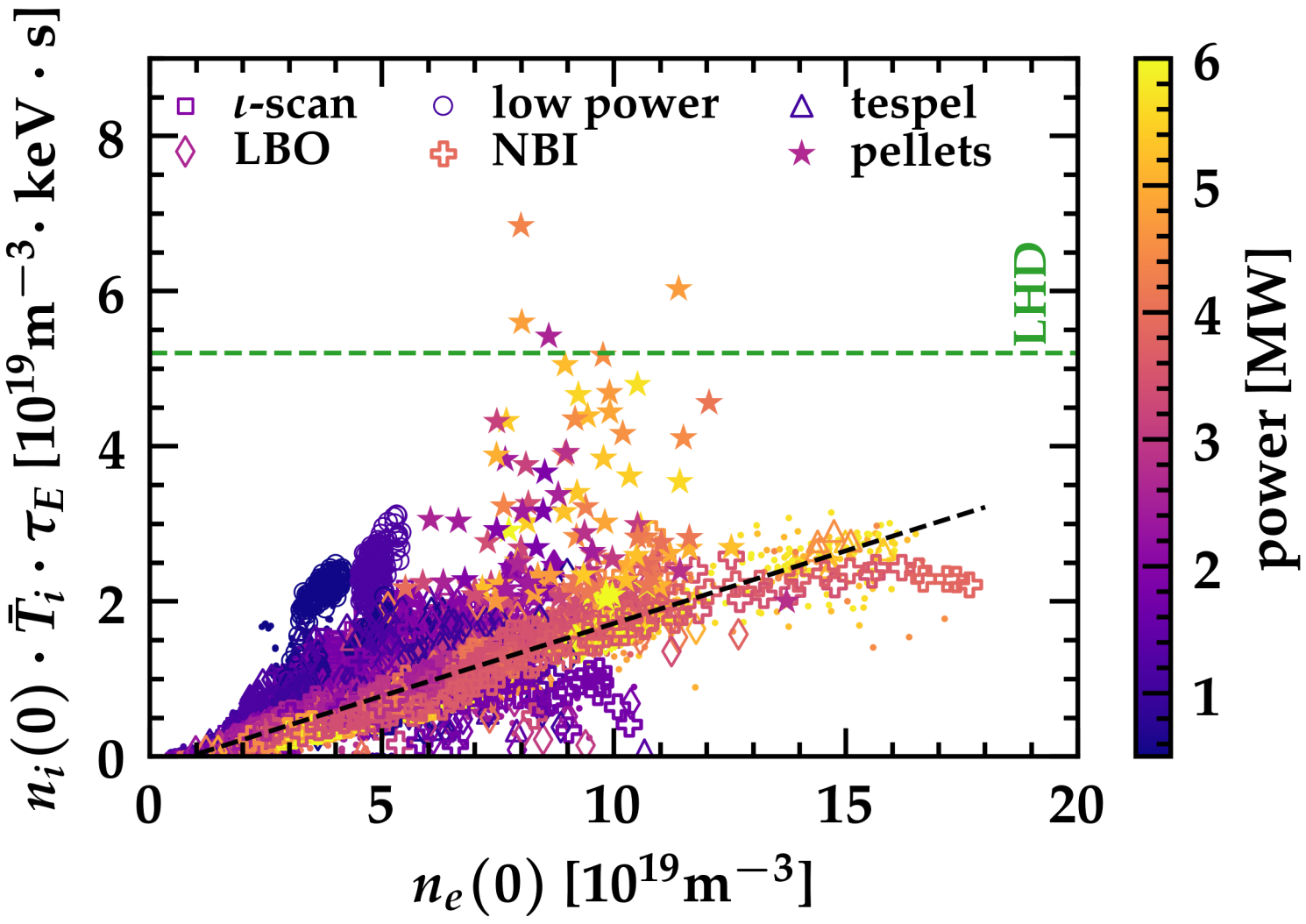

Figure 7. *In this plot, triple products from pellet-fueled discharges have been added (star symbols), clearly illustrating their higher performance.*

Our heuristic description of the "virtuous cycle" induced by pellet injection is illustrated in Figure 8 below.

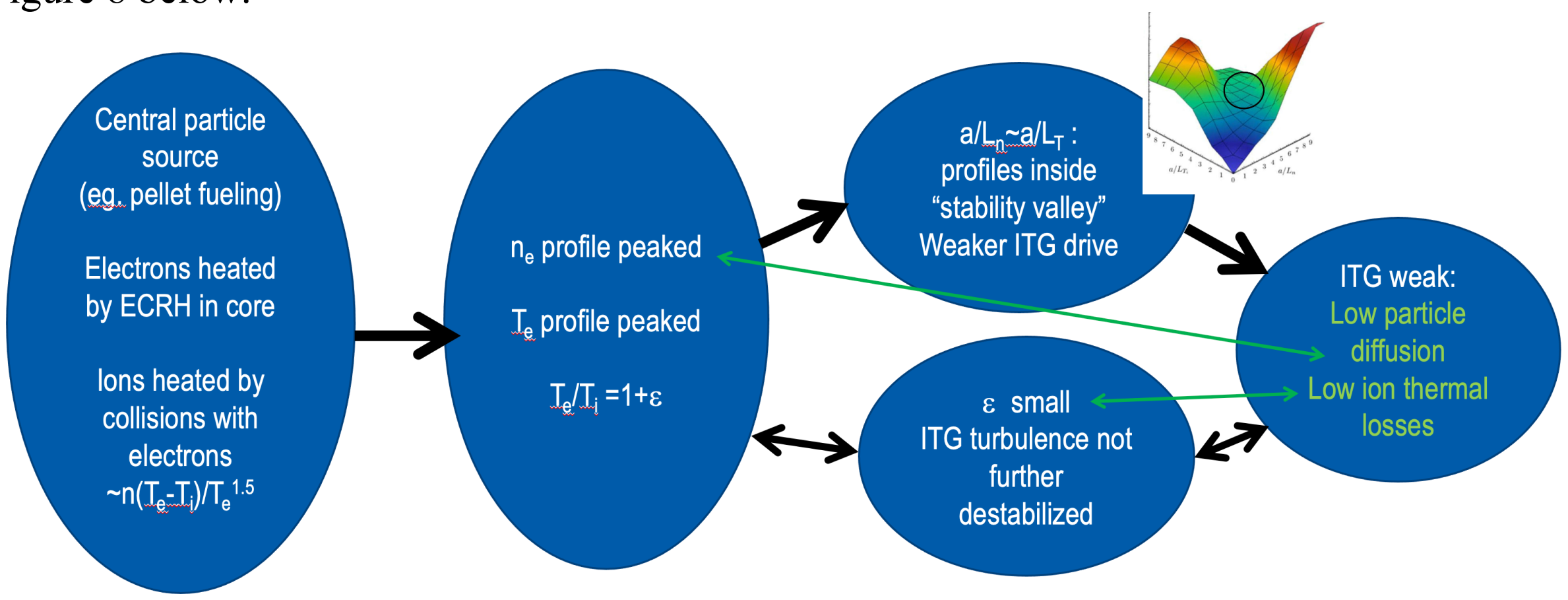

Figure 8. *A "virtuous cycle" can be started by pellet injection. A peaked density profile leads to having profiles residing in the "stability valley", the plasma having a low level of ITG turbulence, and consequently, there are long confinement times of both particles and energy, and a small difference between $T_e$ and $T_i$.*

## 4. Impurity confinement

The virtuous cycle has one downside – it is not entirely virtuous - and the vicious cycle has its advantages:

During the vicious cycle, impurity particle confinement times are low, of order the energy confinement time, and no significant impurity accumulation is seen. For example, using the laser blow-off system of W7-X [42], iron has been injected and is observed spectroscopically to decay with time constants of order 100 ms. There is corroborating evidence that this is due to ITG turbulence: Just as the ITG turbulence itself, the impurity transport is directly influenced by the electron-ion temperature ratio - a larger ratio of Te/Ti leads to larger impurity transport [43]. In turbulent-dominated phases the anomalous contribution to the impurity transport is two orders of magnitude higher than the neoclassical contribution [44].

This transport channel has mainly a diffusive nature [45], and is almost independent of the impurity charge Z [46]. During turbulence-suppressed phases (“virtuous cycle”), we observe that the impurity confinement times are much longer and the impurity content increases [46, 47]. This can be seen from the 2D radiation profiles based on the bolometer measurements viewing the triangular plasma cross section. Those in discharge No. 20180918.45 (described in Fig. 3) obtained by bolometer tomography [48] are shown in Fig. 9. The core emission at t = 3.3 s (where the stored plasma energy $W_p$ reaches its maximum after pellet injection) is more intense than that at t = 5.0s. At the latter time, no pellet effect on the plasma parameters is seen anymore and the plasma stabilizes in a new state. In Fig. 9(c), the radial 1D (flux-surface averaged) emission profiles derived from these 2D distributions are plotted together with those at t=1.5s (before the pellets) and 2.5s (during the pellets) to show the profile evolution.

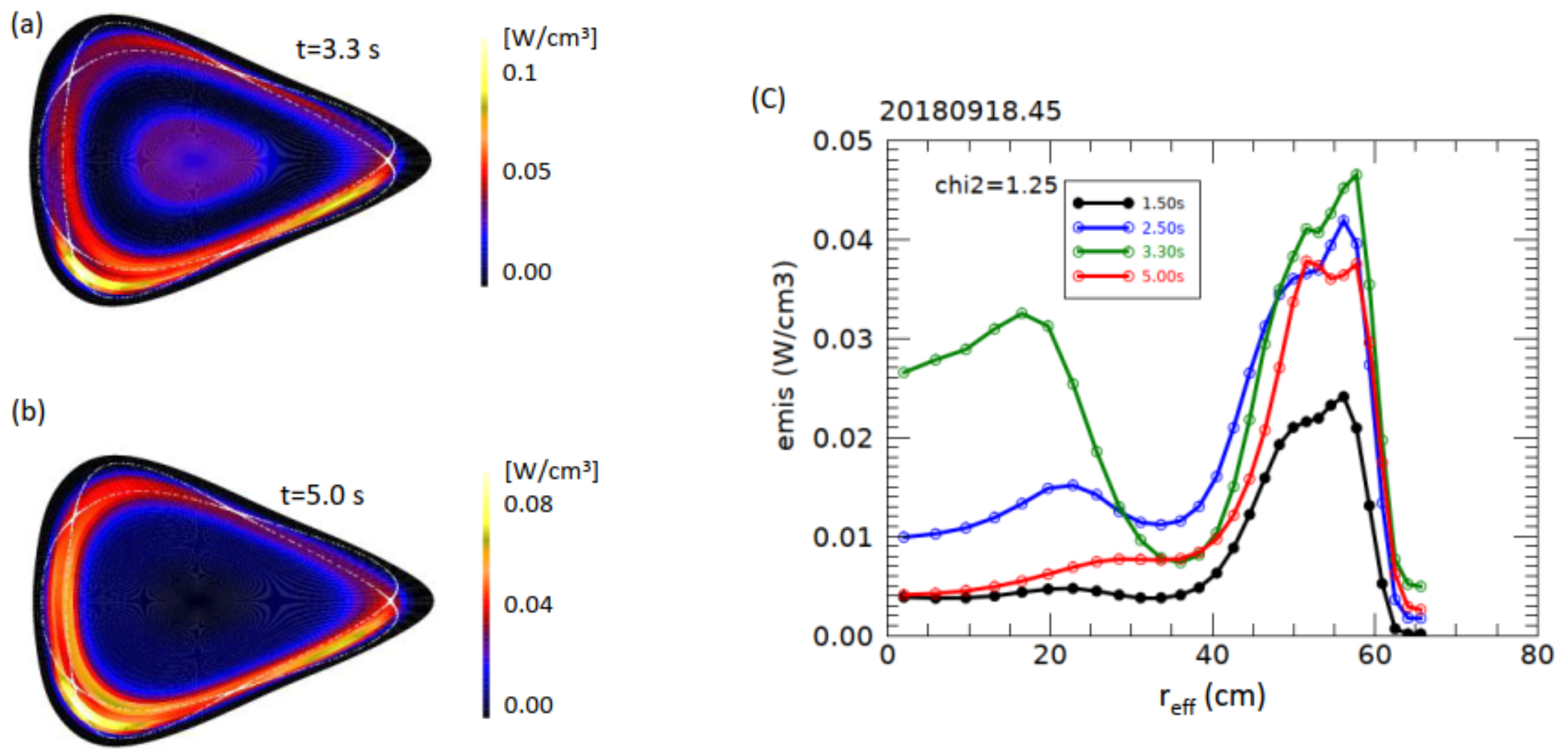

Fig. 9. *2D radiation profiles in the triangular plasma cross section in pellet discharge No. 20180918.45 obtained by bolometer tomography: At t = 3.3 s (a) shortly after pellet injection, where $W_p$ reaches its maximum, there is significant core radiation, consistent with an increased core impurity content, whereas at t = 5.0 s after pellet injections ended, the core radiation has decreased strongly. To the right (c), the radial 1D emission profiles derived from the 2D distributions (flux-area averaged), with the addition of two more time points, t=1.5s (before pellets) and 2.5s (during pellet injection), confirming buildup of core radiation during pellet injection.*

## 5. NBI operation and peaked density profiles

NBI operation also provides a chance to fuel the plasma centrally. High core densities in W7-X were achieved through 3.6 MW of neutral beam injection heating, with no ECRH [39]. These discharges reached core densities of $2\times10^{20}$ $m^{-3}$ with equilibrated ion and electron temperatures at around 1 keV. The addition of 1 MW of ECRH was found to arrest the continual density rise, and raise the temperatures to about 1.5 keV. The density gradient was flattened in the core by the introduction of ECRH, while the density gradient just inside the mid-radius was maintained. Combined heating scenarios will be investigated further in future campaigns, as a path to higher plasma performance.

# 6. General remarks on ITG turbulence

Recent findings from the Large Helical Device (LHD) show that in ITG-dominated discharges hydrogen isotopes readily mix, whereas when electron-scale (trapped-electron mode) turbulence dominates, they do not [50]. It appears that a non-negligible amount of ITG turbulence is beneficial for impurity control as well as for fuel (isotope) exchange and helium exhaust in a stellarator fusion reactor, whereas too much ITG turbulence could potentially clamp the ion temperature below the burn point.

The issues just described present a challenge for W7-X in terms of finding high-confinement (turbulence-suppressed) operational scenarios with benign impurity content – a challenge well-known in tokamaks. These challenges cannot be taken as failures of the computational-optimization approach taken in stellarator optimization, but they do indicate where more research needs to be done. W7-X was optimized for reduced neoclassical losses, but not for reduced turbulent losses. Turbulent transport was not considered during the optimization. It so happens that the benign TEM turbulence is a byproduct of the optimization, and this was realized before operation started [51, 52]. The first numerical results showing promise for targeting benign turbulent transport during the optimization in stellarators appeared about a decade ago [10] and new results give hope that the saturated transport caused by ITG-turbulence can be calculated much faster than with full nonlinear codes [53, 54], which could allow further prospects for turbulence optimization. For W7-X, development of turbulence-suppressed, sustainable discharge scenarios will be a major focus in the upcoming operational phases, ideally combined with benign divertor heat loads and efficient exhaust, hopefully arriving at the demonstration of turbulence-suppressed, high-beta, detached plasmas in steady-state. The major upgrades to the device described in the following will help facilitate such a demonstration.

# 7. Scaling of confinement time with density

It is generally seen – in tokamaks as well as stellarators – that at fixed heating power, higher density leads to longer confinement times. This is reflected in the positive 0.54 exponent for n in the empirical ISS04 scaling [34], developed for stellarators but broadly consistent with tokamak confinement also, $\tau_E = 0.134 a^{2.28} R^{0.64} P_{heat}^{-0.61} n_e^{0.54} B^{0.84} \iota^{0.41}$
It is expressed in terms of minor (a) and major (R) radius, applied heating power $P_{heat}$, plasma density $n_e$, magnetic field strength $B$, rotational transform $\iota$ (iota) at 2/3 of the minor radius, whereas, as discussed in the following, the plasma temperature is an implicit variable.

W7-X "vicious cycle" discharges (with ECRH heating and edge gas particle fueling), are often stationary, allowing a relatively straightforward calculation of confinement times. For such discharges, the density dependence is weaker than predicted by ISS04 although still positive [40]. This scaling is quite consistent with gyro-Bohm transport, as can be shown straightforwardly using a simplified form of the scaling: $\tau_E \propto \left(\frac{n}{P_{heat}}\right)^{\frac{3}{5}} B^{\frac{4}{5}}$ . Using the relationship $P_{heat}\tau_E \propto nT \Rightarrow P_{heat} \propto nT/\tau_E$ to eliminate $P_{heat}$ in favor of plasma parameters $n$ and $T$, one arrives at the gyro-Bohm scaling $\tau_E \propto T^{-\frac{3}{2}} B^2$ with no direct density dependence.
The weaker density dependence found in W7-X "vicious cycle" discharges would appear to indicate that confinement deviates from that expected by gyro-Bohm transport. The analysis of this issue is, however, not straightforward, and a comprehensive treatment will be left to later publications. A short introduction to the complexity involved is given in the following: When the density is increased for discharges of the "vicious cycle" type described in Figure 5, the effect is not simply to lower the temperature(s). The power transferred from the (directly heated) electrons to the ions occurs through Coulomb collisions and is therefore proportional to $n\frac{T_e-T_i}{T_e^{3/2}}$ . Therefore, if the density is increased, the ion temperature will increase, whereas the electron temperature will decrease. So, while one would expect a decrease in electron-scale ETG turbulence, one might simultaneously see an increase in ITG-driven losses. Thus, it is not clear a priori that the turbulent transport losses overall would go down as the density is increased.

The future upgrades described in the next section will allow sustained high densities, and direct ion heating, and electron- and ion temperatures are therefore expected to be much closer to each other. This should allow for a more straightforward study of whether the gyro-Bohm scaling is applicable when scaling to next-step stellarators.

## 8. Device upgrades

The results described here were obtained in OP1.2a and OP1.2b, with an uncooled fine-grain graphite divertor, the Test Divertor Unit (TDU). In order to extend the pulses beyond the limits of its adiabatic temperature-rise limits – which allowed a total heating energy of 200 MJ in one pulse - a water-cooled divertor is now being installed, that will allow up to 18 GJ of heating energy in a single pulse (limited by the adiabatic temperature rise of the finite water reservoir used for cooling). At the time of writing, well over half the divertor modules have been installed, and the project is on track for a completion of in-vessel installation at the end of 2021. The water-cooled divertor will be capable of steady-state heat flux-removal of up to 10 $MW/m^2$, and other components receiving substantial heat loads, such as baffles, and heat shields, are also water cooled in the upcoming phase. However, there will be a number of diagnostic systems that will only attain full steady-state capability on a slightly longer time scale, one of them being the divertor observation system consisting of 10 endoscopes equipped with infrared- as well as visible-light cameras. This system is necessary for safe operation of the divertor, to ensure that the heat flux never exceeds the aforementioned specification. In initial operation, this safe-guarding will occur based on a combination of 7 provisionally improved immersion-tube-based IR observation systems already in operation on OP1.2, as well as 3 steady-state capable IR endoscopes. The transition to 10 fully steady-state capable IR endoscopes will stretch into 2024.
In each of the 10 subdivertor spaces, a cryopump will be installed, which will increase the effective exhaust rate from appr. 30 000 l/s to appr. 130 000 l/s, a factor of 4.3. This will further ensure efficient exhaust of particles.

An important system, in particular for turbulence suppression and high performance, will be the continuous pellet fueling system (CPTS) [55], which is projected to be operational at or shortly after first plasma in fall 2022. This system will allow for high and sustainable central fueling with hydrogen and deuterium ice pellets with injection velocities up to 1000 m/s ie. more than a factor of three higher than those of the blower-gun system that was operating in OP1.2, whose pellet injection speed was limited to 300 m/s. The CPTS system is being developed in a joint venture led by US partners at Oak Ridge National Laboratory and Princeton Plasma Physics Laboratory, with substantial resources from NIFS in Japan and also the W7-X home team. High hydrogen ice production rates have been achieved in prototype testing, exceeding $10^{22}$ H atoms/sec for tens of minutes, and up to twice that for up to three minutes. The necessary fueling rate to allow for sustained density peaking (and thereby sustained reduction of ITG turbulence) is not precisely known, but it is not unlikely that this system will be capable of sustaining the density peaking for the full duration of a long-pulse discharge in W7-X.

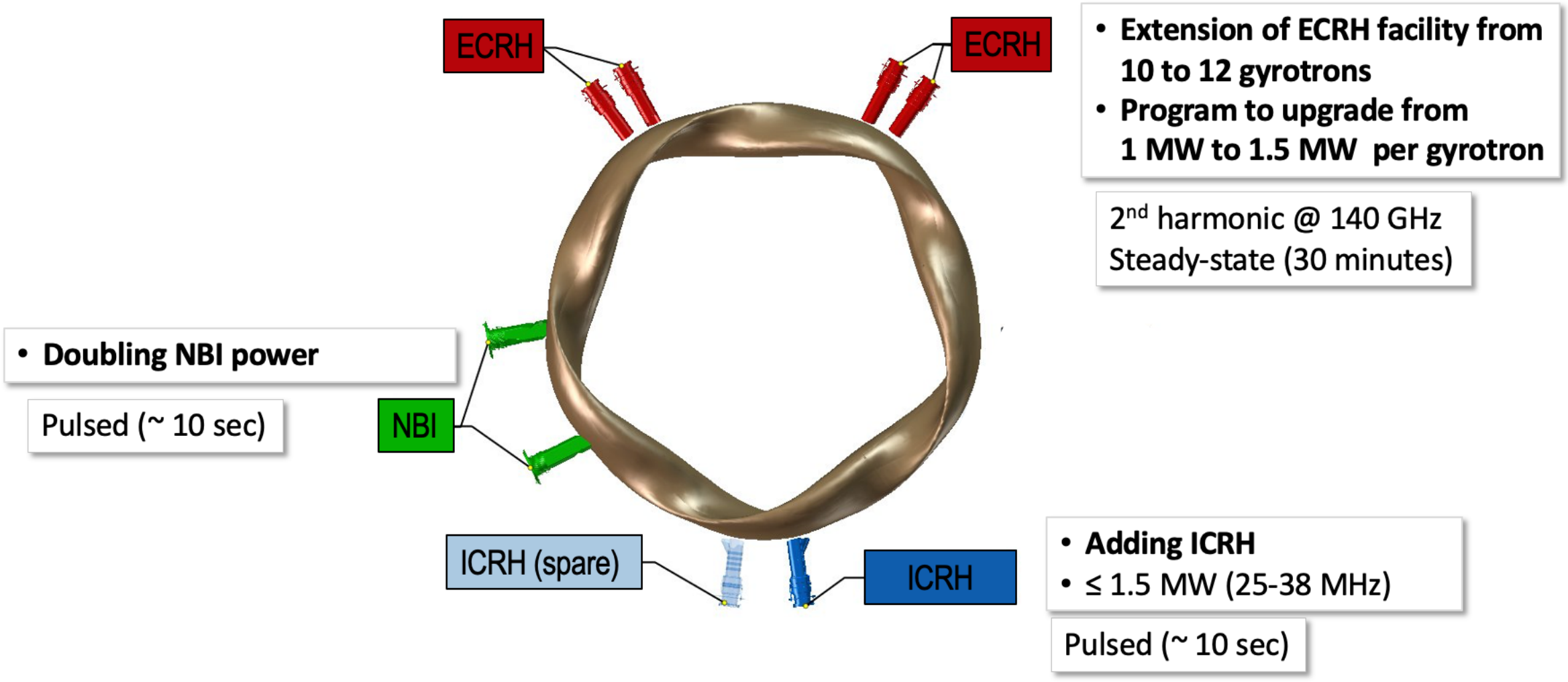

Figure 10. *The layout and upgraded capabilities of heating systems for the upcoming operation phase*

The heating systems will also see major upgrades, Fig. 10. An increase from 10 to 12 gyrotrons will give a modest increase in the near-steady-state ECRH heating capabilities, and 1.5 MW gyrotrons are being developed to replace the present ones that are limited to maximally 1 MW per gyrotron. The NBI heating system will be doubled from 3.6 to 7.2 MW for hydrogen operation and up to 10 MW operated with deuterium– this system is currently limited to operate for 10 seconds at a time. An ICRH system will also be installed, adding 1.5 MW of heating at an adjustable frequency in the range from 25 to 38 MHz [56]. This system will allow plasma startup at lower field strengths as well as dedicated studies of the confinement of high-energy ions.

## 9. Summary

The results obtained during preliminary operation of W7-X have been very encouraging. The island divertor works remarkably well, and the neoclassical optimization is successful. Impurities are not a problem during turbulence-dominated discharges, and are well screened from the core plasma by the W7-X divertor. However, impurity injection experiments during turbulence-suppressed phases show that impurity accumulation could be a challenge for long-pulse, high-performance discharges in the upcoming experimental campaigns. Besides the references already given, readers are also referred to other recent W7-X publications [57-61] for more examples of recent results and more detailed descriptions.

## 10. Acknowledgements

This work has been carried out within the framework of the EUROfusion Consortium and has received funding from the Euratom research and training programme 2014-2018 and 2019-2020 under grant agreement No 633053. The views and opinions expressed herein do not necessarily reflect those of the European Commission.

## Appendix 1: W7-X Team List

T. Sunn Pedersen [1, 2], I. Abramovic [3], P. Agostinetti [4], M. Agredano Torres [1], S. Äkäslompolo [1], J. Alcuson Belloso [1], P. Aleynikov [1], K. Aleynikova [1], M. Alhashimi [1], A. Ali [1], N. Allen [5], A. Alonso [6], G. Anda [7], T. Andreeva [1], C. Angioni [8], A. Arkhipov [8], A. Arnold [1], W. Asad [8], E. Ascasibar [6], M.-H. Aumeunier [9], K. Avramidis [10], E. Aymerich [11], S.-G. Baek [3], J. Bähner [1], A. Baillod [12], M. Balden [1], M. Balden [8], J. Baldzuhn [1], S. Ballinger [3], M. Banduch [1], S. Bannmann [1], A. Banon Navarro [8], A. Bañón Navarro [1], T. Barbui [13], C. Beidler [1], C Belafdil [9], A. Bencze [7], A. Benndorf [1], M. Beurskens [1], C. Biedermann [1], O. Biletskyi [14], B. Blackwell [15], M. Blatzheim [1], T. Bluhm [1], D. Böckenhoff [1], G. Bongiovi [16], M. Borchardt [1], D. Borodin [17], J. Boscary [8], H. Bosch [1, 18], T. Bosmann [19], B. Böswirth [8], L. Böttger [1], A. Bottino [8], S. Bozhenkov [1], R. Brakel [1], C. Brandt [1], T. Bräuer [1], H. Braune [1], S. Brezinsek [17], K. Brunner [1], S. Buller [1], R. Burhenn [1], R. Bussiahn [1], B. Buttenschön [1], A. Buzás [7], V. Bykov [1], I. Calvo [6], K. Camacho Mata [1], I. Caminal [20], B. Cannas [11], A. Cappa [6], A. Carls [1], F. Carovani [1], M. Carr [21], D. Carralero [6], B. Carvalho [22], J. Casas [20], D. Castano-Bardawil [17], F. Castejon [6], N. Chaudhary [1], I. Chelis [23], A. Chomiczewska [24], J.W. Coenen [17, 13], M. Cole [1], F. Cordella [25], Y. Corre [9], K. Crombe [26], G. Cseh [7], B. Csillag [7], H. Damm [1], C. Day [10], M. de Baar [27], E. De la Cal [6], S. Degenkolbe [1], A. Demby [13], S. Denk [3], C. Dhard [1], A. Di Siena [8, 28], A. Dinklage [1, 2], T. Dittmar [17], M. Dreval [14], M. Drevlak [1], P. Drewelow [1], P. Drews [17], D. Dunai [7], E. Edlund [3], F. Effenberg [29], G. Ehrke [1], M. Endler [1], D. A. Ennis [5], F.J. Escoto [6], T. Estrada [6], E. Fable [8], N. Fahrenkamp [1], A. Fanni [11], J. Faustin [1], J. Fellinger [1], Y. Feng [1], W. Figacz [24], E. Flom [13], O. Ford [1], T. Fornal [24], H. Frerichs [13], S. Freundt [1], G. Fuchert [1], M. Fukuyama [30], F. Füllenbach [1], G. Gantenbein [10], Y. Gao [1], K. Garcia [13], J. M. García Regaña [6], I. García-Cortés [6], J. Gaspar [31], D. A. Gates[29], J. Geiger [1], B. Geiger [13], L. Giudicotti [32], A. González [6], A. Goriaev [26, 33], D. Gradic [1], M. Grahl [1], J. P. Graves [12], J. Green [13], E. Grelier [9], H. Greuner [8], S. Groß [1], H. Grote [1], M. Groth [34], M. Gruca [24], O. Grulke [1,41], M. Grün [1], J. Guerrero Arnaiz [1], S. Günter [8], V. Haak [1], M. Haas [1], P. Hacker [1], A. Hakola [35], A. Hallenbert [1], K. Hammond [29], X. Han [17, 36], S.K. Hansen [3], J.H. Harris [37], H. Hartfuß [1], D. Hartmann [1], D. Hathiramani [1], R. Hatzky [8], J. Hawke [38], S. Hegedus [7], B. Hein [8], B. Heinemann [8], P. Helander [1, 2], S. Henneberg [1], U. Hergenhahn [8, 57], C. Hidalgo [6], F. Hindenlang [8], M. Hirsch [1], U. Höfel [1], K. P. Hollfeld [17], A. Holtz [1], D. Hopf [8], D. Höschen [17], M. Houry [9], J. Howard [19], X. Huang [39], M. Hubeny [17], S. Hudson [29], K. Ida [39], Y. Igitkhanov [10],

V. Igochine [8], S. Illy [10], C. Ionita-Schrittwieser [40], M. Isobe [39], M. Jabłczyńska [24], S. Jablonski [24], B. Jagielski [1], M. Jakubowski [1], A. Jansen van Vuuren [1], J. Jelonnek [10], F. Jenko [8], F. Jenko [8], T. Jensen [41], H. Jenzsch [1], P. Junghanns [8], J. Kaczmarczyk [24], J. Kallmeyer [1], U. Kamionka [1], M. Kandler [8], S. Kasilov [42], Y. Kazakov [26], D. Kennedy [1], A. Kharwandikar [1], M. Khokhlov [1], C. Kiefer [8], C. Killer [1], A. Kirschner [17], R. Kleiber [1], T. Klinger [1, 2], S. Klose [1], J. Knauer [1], A. Knieps [17], F. Köchl [43], G. Kocsis [7], Ya.I. Kolesnichenko [44], A. Könies [1], R. König [1], J. Kontula [34], P. Kornejew [1], J. Koschinsky [1], M. M. Kozulia [14], A. Krämer-Flecken [17], R. Krampitz [1], M. Krause [1], N. Krawczyk [24], T. Kremeyer [1], L. Krier [10], D. M. Kriete [5], M. Krychowiak [1], I. Ksiazek [45], M. Kubkowska [24], M. Kuczynski [1], G. Kühner [1], A. Kumar [15], T. Kurki-Suonio [34], S. Kwak [1], M. Landreman [46], P. T. Lang [8], A. Langenberg [1], H.P. Laqua [1, 2], H. Laqua [1], R. Laube [1], S. Lazerson [1], M. Lewerentz [1], C. Li [17], Y. Liang [17], Ch. Linsmeier [17], J. Lion [1], A. Litnovsky [17, 58], S. Liu [36], J. Lobsien [1], J. Loizu [12], J. Lore [37], A. Lorenz [1], U. Losada [6], F. Louche [26], R. Lunsford [29], V. Lutsenko [44], M. Machielsen [12], F. Mackel [8], J. Maisano-Brown [3], O. Maj [8], D. Makowski [47], G. Manduchi [48], E. Maragkoudakis [6], O. Marchuk [17], S. Marsen [1], E. Martines [4], J. Martinez-Fernandez [6], M. Marushchenko [1], S. Masuzaki [39], D. Maurer [5], M. Mayer [8], K. J. McCarthy [6], O. Mccormack [4], P. McNeely [1], H. Meister [8], B. Mendelevitch [8], S. Mendes [1], A. Merlo [1], A. Messian [26], A. Mielczarek [47], O. Mishchenko [1], B. Missal [1], R. Mitteau [9], V. E. Moiseenko [14], A. Mollen [1], V. Moncada [9], T. Mönnich [1], T. Morisaki [39], D. Moseev [1], G. Motojima [39], S. Mulas [6], M. Mulsow [1], M. Nagel [1], D. Naujoks [1], V. Naulin [41], T. Neelis [19], H. Neilson [29], R. Neu [8], O. Neubauer [17], U. Neuner [1], D. Nicolai [17], S. K. Nielsen [41], H. Niemann [1], T. Nishiza [1], T. Nishizawa [1], T. Nishizawa [8], C. Nührenberg [1], R. Ochoukov [8], J. Oelmann [17], G. Offermanns [17], K. Ogawa [39], S. Okamura [39], J. Ölmanns [17], J. Ongena [26], J. Oosterbeek [1], M. Otte [1], N. Pablant [29], N. Panadero Alvarez [6], N. Panadero Alvarez [6], A. Pandey [1], E. Pasch [1], R. Pavlichenko [14], A. Pavone [1], E. Pawelec [45], G. Pechstein [1], G. Pelka [24], V. Perseo [1], B. Peterson [39], D. Pilopp [1], S. Pingel [1], F. Pisano [11], B. Plöckl [8], G. Plunk [1], P. Pölöskei [1], B. Pompe [2], A. Popov [49], M. Porkolab [3], J. Proll [19], M.J. Pueschel [19, 27], M.-E. Puiatti [50], A. Puig Sitjes [1], F. Purps [1], K. Rahbarnia [1], M. Rasiński [17], J. Rasmussen [41], A. Reiman [29], F. Reimold [1], M. Reisner [8], D. Reiter [17], M. Richou [9], R. Riedl [8], J. Riemann [1], K. Riße [1], G. Roberg-Clark [1], V. Rohde [8], J. Romazanov [17], D. Rondeshagen [1], P. Rong [1], L. Rudischhauser [1], T. Rummel [1], K. Rummel [1], A. Runov [1], N. Rust [1], L. Ryc [24], P. Salembier [20], M. Salewski [41], E. Sanchez [6], S. Satake [39], G. Satheeswaran [17], J. Schacht [1], E. Scharff [1], F. Schauer [8], J. Schilling [1], G. Schlisio [1], K. Schmid [8], J. Schmitt [5], O. Schmitz [13], W. Schneider [1], M. Schneider [1], P. Schneider [8], R. Schrittwieser [40], T. Schröder [1], M. Schröder [1], R. Schroeder [1], B. Schweer [26], D. Schwörer [1], E. Scott [1], E. Scott [8], B. Shanahan [1], G. Sias [11], P. Sichta [29], M. Singer [1], P. Sinha [29], S. Sipliä [34], C. Slaby [1], M. Sleczka [51], H. Smith [1], J. Smoniewski [52], E. Sonnendrücker [8], M. Spolaore [4], A. Spring [1], R. Stadler [8], D. Stańczak [24], T. Stange [1], I. Stepanov [26], L. Stephey [13], J. Stober [8], U. Stroth [8, 53], E. Strumberger [8], C. Suzuki [39], Y. Suzuki [39], J. Svensson [1], T. Szabolics [7], T. Szepesi [7], M. Szücs [7], F. L. Tabarés [6], N. Tamura [39], A. Tancetti [41], C. Tantos [10], J. Terry [3], H. Thienpondt [6], H. Thomsen [1], M. Thumm [10], J.M. Travere [9], P. Traverso [5], J. Tretter [8], E. Trier [8], H. Trimino Mora [1], T. Tsujimura [39], Y. Turkin [1], A. Tykhyi [44], B. Unterberg [17], P. van Eeten [1], B.Ph. van Milligen [6], M. van Schoor [26], L. Vano [1], S. Varoutis [10], M. Vecsei [7], L. Vela [56], J. L. Velasco [6], M. Vervier [17], N. Vianello [48], H. Viebke [1], R. Vilbrandt [1], G. Vogel [8], N. Vogt [1], C. Volkhausen [1], A. von Stechow [1], F. Wagner [1], E. Wang [17], H. Wang [54], F. Warmer [1], T. Wauters [26], L. Wegener [1], T. Wegner [1], G. Weir [1], U. Wenzel [1], A. White [3], F. Wilde [1], F. Wilms [1], T. Windisch [1], M. Winkler [1], A. Winter [1], V. Winters [1], R. Wolf [1, 18], A. M. Wright [29], G. A. Wurden [38], P. Xanthopoulos [1], S. Xu [17], H. Yamada [55], H. Yamaguchi [39], M. Yokoyama [39], M. Yoshinuma [39], Q. Yu [8], M. Zamanov [14], M. Zanini [1], M. Zarnstorff [29], D. Zhang [1], S. Zhou [17], J. Zhu [1], C. Zhu [29], M. Zilker [8], A. Zocco [1], H. Zohm [8], S. Zoletnik [7], L. Zsuga [7]

1 Max-Planck Institute for Plasma Physics, Wendelsteinstrasse 1, 17491 Greifswald, Germany

2 University of Greifswald, Greifswald, Germany
3 Massachusetts Institute of Technology, 77 Massachusetts Ave, Cambridge, MA 02139, United States of America
4 Consorzio RFX, Corso Stati Uniti, 4 – 35127 Padova, Italy
5 Auburn University, Auburn, AL 36849, United States of America
6 CIEMAT, Avenida Complutense, 40, 28040 Madrid, Spain
7 Centre for Energy Research, Konkoly-Thege út 29-33, 1121 Budapest, Hungary
8 Max-Planck-Institute for Plasma Physics, Boltzmannstraße 2, 85748 Garching bei München, Germany
9 CEA Cadarache, 13115 Saint-Paul-lez-Durance, France
10 Karlsruhe Institute of Technology, Kaiserstr. 12, 76131 Karlsruhe, Germany
11 University of Cagliari, Via Università, 40, 09124 Cagliari, Italy
12 École Polytechnique Fédérale de Lausanne, Swiss Plasma Center, CH-1015 Lausanne, Switzerland
13 University of Wisconsin–Madison, Engineering Drive, Madison, Wisconsin 53706, United States of America
14 Institute of Plasma Physics, National Science Center "Kharkiv Institute of Physics and Technology", Kharkiv, Ukraine
15 The Australian National University, Acton ACT 2601, Canberra, Australia
16 University of Palermo, Department of Engineering, Viale delle Scienze, Edificio 6, Palermo, 90128, Italy
17 Forschungszentrum Jülich GmbH, Institut für Energie- und Klimaforschung – Plasmaphysik, 52425 Jülich, Germany
18 Technical University of Berlin, Strasse des 17. Juni 135, 10623 Berlin, Germany
19 Eindhoven University of Technology, 5600 MB Eindhoven, The Netherlands
20 Universitat Politècnica de Catalunya. BarcelonaTech, C. Jordi Girona, 31, 08034 Barcelona, Spain.
21 Culham Centre for Fusion Energy, Abingdon OX14 3EB, United Kingdom
22 Instituto de Plasmas e Fusao Nuclear, Av. Rovisco Pais, 1049-001 Lisboa, Portugal
23 Department of Physics, National and Kapodistrian University of Athens, 157 84 Athens, Greece
24 Institute of Plasma Physics and Laser Microfusion, 23 Hery Str., 01-497 Warsaw, Poland
25 ENEA - Centro Ricerche Frascati, Via Enrico Fermi, 45, 00044 Frascati RM, Italy
26 Laboratory for Plasma Physics, LPP-ERM/KMS, TEC Partner, B-1000 Brussels, Belgium
27 Dutch Institute for Fundamental Energy Research, PO Box 6336, 5600 HH Eindhoven, Netherlands
28 University of Texas, Austin, TX, USA
29 Princeton Plasma Physics Laboratory, Princeton, NJ 08543, USA
30 Kyushu University, 744 Motooka Nishi-ku, Fukuoka 819-0395, Japan.
31 Aix-Marseille University, Jardin du Pharo, 58 Boulevard Charles Livon, 13007, Marseille, France
32 Padova University, Department of Physics and Astronomy, Via Marzolo 8 , 35131 Padova, Italy
33 Department of Applied Physics, Ghent University, Sint-Pietersnieuwstraat 41 B4, 9000 Ghent, Belgium
34 Aalto University, 02150 Espoo, Finland
35 VTT Technical Research Centre of Finland Ltd., PO Box 1000, FI-02044 VTT, Finland
36 Institute of Plasma Physics, Chinese Academy of Sciences, 230031 Hefei, Anhui, People's Republic of China
37 Oak Ridge National Laboratory, 1 Bethel Valley Rd, Oak Ridge, TN 37830, United States of America

38 Los Alamos National Laboratory, NM 87545 United States of America
39 National Institute for Fusion Science, National Institutes of Natural Sciences, 322-6 Oroshi-cho, Toki, Gifu Prefecture 509-5292, Japan
40 Institute for Ion Physics and Applied Physics, University of Innsbruck, Innsbruck, Austria
41 Technical University of Denmark, Department of Physics, Anker Engelunds Vej, 2800 Kgs Lyngby, Denmark
42 Graz University of Technology, Rechbauerstraße 12, 8010 GRAZ, AUSTRIA
43 Austrian Academy of Science, Doktor-Ignaz-Seipel-Platz 2, 1010 Wien, Austria
44 Institute for Nuclear Research, prospekt Nauky 47, Kyiv 03028, Ukraine
45 University of Opole, plac Kopernika 11a, 45-001 Opole, Poland
46 University of Maryland, Paint Branch Drive, College Park, MA 20742, United States of America
47 Lodz University of Technology, Department of Microelectronics and Computer Science Wolczanska 221/223, 90-924 Lodz, Poland
48 Consiglio Nazionale delle Ricerche, Piazzale Aldo Moro, 7, 00185 Roma, Italy.
49 Ioffe Physical-Technical Institute of the Russian Academy of Sciences, 26 Politekhnicheskaya, St Petersburg 194021, Russian Federation
50 Istituto di Fisica del Plasma Piero Caldirola, Via Roberto Cozzi, 53, 20125 Milano, Italy
51 University of Szczecin, 70-453, aleja Papieża Jana Pawła II 22A, Szczecin, Poland
52 Lawrence University, 711 E Boldt Way, Appleton, WI 54911, USA.
53 Physik-Department E28, Technische Universität München, 85747 Garching, Germany
54 Yale University, New Haven, Connecticut 06520, USA.
55 University of Tokyo, 5-1-5 Kashiwanoha, Kashiwa, Chhiab 277-0882 Japan
56 Universidad Carlos III de Madrid, Av. de la Universidad, 30, Madrid, Spain
57 Fritz-Haber-Institut der Max-Planck-Gesellschaft, 14195 Berlin, Germany
58 National Research Nuclear University MEPhI, 115409 Moscow, Russian Federation